\documentclass[10pt,conference]{IEEEtran}

\usepackage{amsmath,amsfonts}
\usepackage[ruled,vlined]{algorithm2e}
\usepackage{hyperref}
\usepackage{graphicx}
\usepackage{array}
\usepackage{amssymb}
\usepackage{graphicx}
\usepackage{color}
\usepackage[english]{babel}

\usepackage{tabularx}

\begin{document}
\title{On the Modeling of Reliability in Extreme Edge Computing Systems}
\author{Mhd Saria Allahham$^{1}$,  Amr Mohamed$^2$, Aiman Erbad$^3$ and Hossam Hassanein$^1$\\
		$^1$School of Computing, Queen's University, Kingston, ON, Canada\\ 
		$^2$College of Engineering, Qatar University, Qatar. \\
		$^3$College of Science and Engineering, Hamad Bin Khalifa University, Qatar\\ 
		Email: 20msa7@queensu.ca, amrm@qu.edu.qa, aerbad@ieee.org, and hossam@cs.queensu.ca 
}
\maketitle
\begin{abstract}
Extreme edge computing (EEC) refers to the endmost part of edge computing wherein computational tasks and edge services are deployed only on extreme edge devices (EEDs). EEDs are consumer or user-owned devices that offer computational resources, which may consist of wearable devices, personal mobile devices, drones, etc. Such devices are opportunistically or naturally present within the proximity of other user devices. Hence, utilizing EEDs to deploy edge services or perform computational tasks fulfills the promise of edge computing of bringing the services and computation as close as possible to the end-users. However, the lack of knowledge and control over the EEDs computational resources raises a red flag, since executing the computational tasks successfully becomes doubtful. To this end, we aim to study the EEDs randomness from the computational perspective, and how reliable is an EED in terms of executing the tasks on time. Specifically, we provide a reliability model for the EEDs that takes into account the probabilistic nature of the availability of the EEDs' computational resources. Moreover, we study the reliability of executing different types of computational tasks in EEC systems that are distributed across the EEDs. Lastly, we carry out experimental results to analyze the EEDs and the EEC systems' reliability behavior in different scenarios.
\end{abstract}

\section{Introduction}

Edge computing is a paradigm that aims at bringing streaming services such as extended reality (XR), online gaming, content delivery services, etc., and computational services such as distributed Machine Learning (ML) training and inference \cite{FL,baccour2021pervasive}, closer to the end user, without the need for the cloud \cite{ranaweera2021survey}. This is due to the fact that utilizing the cloud resources suffers from a huge latency overhead, besides that such computational tasks and services are time-restricted and delay-sensitive, respectively. In edge computing, tasks are offloaded from resource-limited edge devices to enterprise-owned devices such as edge servers, which are usually located at the Radio Access Network (RAN) sites \cite{edge_comp_ran, tran2017collaborative}. Conversely, Extreme Edge Computing (EEC), which is the endmost part of the edge computing continuum, aims at offloading the tasks only to consumer and user-owned devices that offer computational resources rather than the edge servers. Examples of such devices can be smart wearable devices, personal mobile devices, and smart home appliances, drones, etc. These Extreme Edge Devices (EEDs) are abundant and naturally the nearest part of the edge to the end users, and may possess better connectivity with the end users than the RANs. In fact, utilizing such devices for computation or services allows for achieving the low-latency premise of edge computing. 
However, this comes at the cost of encountering the challenge of the unpredictable behaviour of such devices in terms of the availability of their computational resources. In fact, it is impractical to have complete knowledge about the computational behaviour and the local tasks that are being executed by the EEDs, whether for privacy or other concerns. Moreover, the tasks or services at the edge may have computational demands that change with time in a deterministic or stochastic manner. Indeed, the EED's capability to allocate the required computational resources, finish the offloaded tasks on time, or guarantee an uninterruptible edge service becomes questionable. 
Therefore, it is essential to study the reliability of the EEDs from the computational perspective, and more specifically, the ability of such devices to finish computational tasks on time and provide ceaseless and steady edge service, given that there exists uncertainty in the tasks' demands or the EEDs' computational resources. Moreover, since computational tasks or services in an EEC system can be distributed across the available EEDs in the system as sub-tasks or sub-services, the studying of the reliability of an EEC system in terms of executing the task successfully or providing a seamless service in a distributed manner is of great importance.

A plethora of works has addressed the reliability in edge computing from different perspectives. For instance, Wu et al. \cite{wu2020accelerating} have proposed a client selection algorithm that takes into account the reliability of the clients to improve the training process in Federated Learning. The reliability is simply assumed as the probability that a client will not be dropped out during the training, where the probability is sampled from a Gaussian distribution. Whereas the works in \cite{liu2021reliability, li2019providing} have addressed the reliability of the containers and virtual machines (VMs) at the edge servers, considering that there exist software failures when initiating instances or admitting service requests. To generalize the reliability model, the works in \cite{hou2020latency} and \cite{hou2020reliable} have considered a joint model for the computation and communication reliability for drones and smart vehicles, respectively. In both works, the computation and communication reliability model are represented by an exponential distribution model. From a different perspective, the authors in \cite{liu2019fast} have considered the reliability considering the transient failure (i.e., the random hardware components failure of edge devices). The hardware reliability of the devices was assumed to follow an exponential distribution.

Even though the reliability has been addressed in the literature from various perspectives, the computational perspective of the EEDs and how it is affected by the task computational demand and the utilization of available resources has been disregarded. Moreover, all of the previous works assumed complete knowledge about the exact allocated computational resources of the EEDs, which may not be always the case due to the lack of control over these devices, and the lack of information about the other local tasks that are already being executed by the EED. To this end, we opt to model and study the EEDs' computational reliability, and the ability of the EEDs to execute the computational tasks on time considering probabilistic task demands and available computational resources. The scope of this work is narrowed to focus only on the reliability of computational task execution.
The contributions of this work can be summarized as follows: 
\begin{enumerate}
    \item We provide a statistical model for the EEDs reliability considering different behaviours of task demands and the availability of computational resources for the tasks.
    \item We elaborate on the case of probabilistic task demand and computational resources, and derive a closed-form expression for the EEDs reliability. 
    \item We study and model the reliability of different types of EEC systems according to different computational task types.
\end{enumerate}

The rest of the paper is organized as follows: We present the EEDs reliability model in \ref{sec:EED_rel}, while section \ref{sec:Sys_rel} present the EEC systems reliability. We show the simulation results in Section \ref{sec:sim}. We then discuss the work and future direction in \ref{sec:discussion} before we conclude in Section \ref{sec:conc}.

\section{Modeling of Extreme Edge Nodes Reliability}\label{sec:EED_rel}

Let $T$ be a random variable (RV) that follows a distribution $f_T$, where $f_T$ represents the Probability Density Function (PDF), $T$ is the time taken for an EED to finish its task. Therein, we define the reliability of the EED as the probability of finishing a task before the deadline $t$ as follows:
\begin{equation}
R(t) := F(t) = P(T \leq t) = \int_0^t f_T(x) dx,
\end{equation}
where $F(t)$ is the Cumulative Distribution Function (CDF). 

\begin{figure}
    \centering
    \includegraphics[scale=0.487]{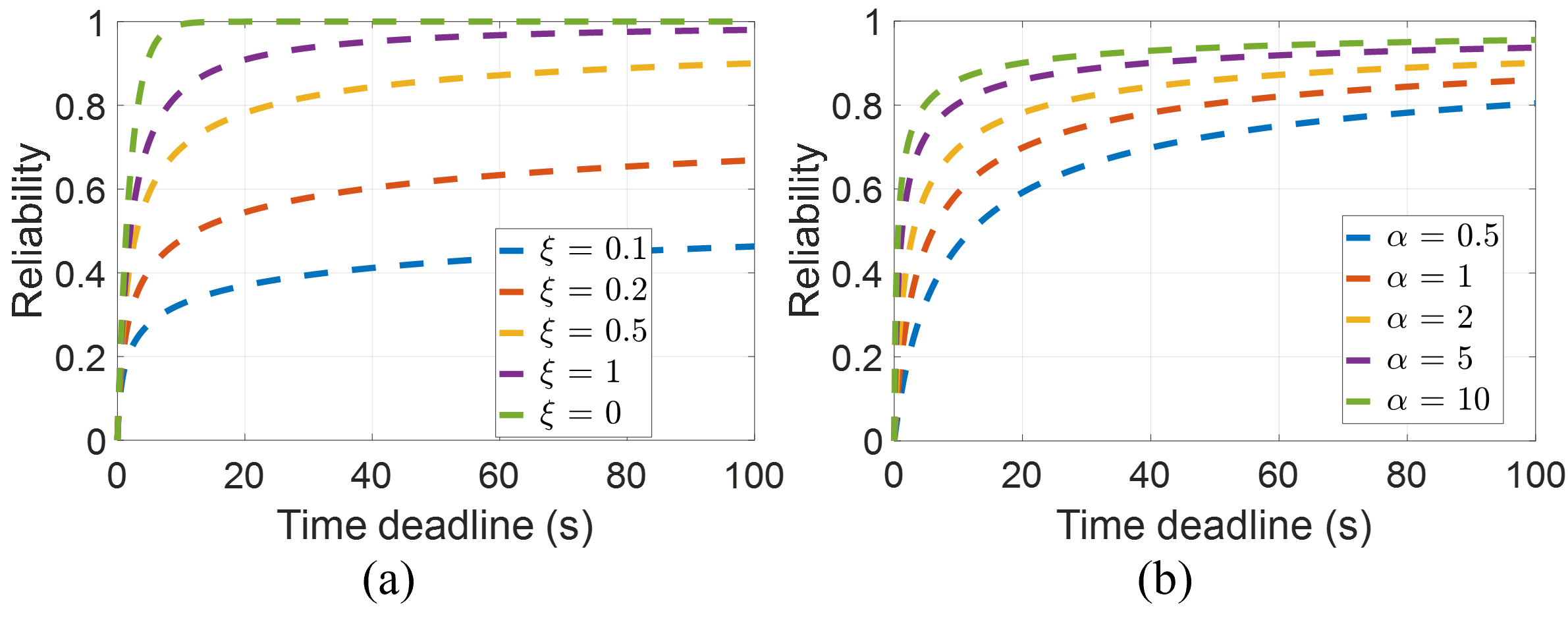}
    \caption{The Generalized Pareto Distribution with different (a) shapes at $\alpha = 2$  (b) scales at $\xi = 2$}
    \label{fig:GDP}
\end{figure}

As for the time distribution $f_T$, while several works have considered an exponential distribution, Balter in \cite{harchol2013performance} claimed that the computational task times in almost all systems follow a Type 1 Pareto distribution. However, considering one distribution over another might not generalize every EEC system, since the computational capacities of EEDs are heterogeneous, and there is a huge variety of tasks at the edge, each of which has different demands and characteristics. Therefore, we opt to use a more generalized model, namely, the Generalized Pareto distribution (GPD) \cite{GDP}. The GPD provides more degrees of freedom in modeling as it is a generalization of exponential distribution and all types of Pareto distributions. Therein, we define the reliability of the EEDs as the CDF of the GPD, and is defined as follows:
\begin{equation}
F(t;\alpha,\xi) = \begin{cases}
1- (1+\frac{t \alpha}{\xi})^{-\xi }  & \xi^{-1} > 0, \\
1- e^{-\alpha t}& \xi^{-1} =0, 
\end{cases}
\end{equation}
where $\xi$ and $\alpha$ are non-negative distribution parameters. As it can be seen from Fig. \ref{fig:GDP}, the parameter $\xi$ represents the asymptotic tail behaviour of the distribution, whereas the parameter $\alpha$, it represents how the reliability increases over time, or the rate at which the reliability approach the asymptotic behaviour. The asymptotic term represents the behaviour where the increase in reliability over time become insignificant, such that for $t_2 >> t_1$, $R(t_2) \approx R(t_1)$. The distribution tail conveys that for some tasks, the EED may take huge amount of time to execute the task, or may not execute it at all, which results in making the EED less reliable. As such, each EED can be characterized by a unique constant $\xi$, where the asymptotic tail can indicate the maximum attainable reliability from that EED for a specific time horizon. 



The $k^{th}$ statistical moment of the GPD is only defined for $\xi > k$. Therein, the first moment of the GPD is given by:
\begin{equation}
\mathbb{E}[T] = \begin{cases} 
 \frac{1}{\alpha} & \xi = 0 \\
 \\
\frac{1}{\alpha ( 1  - \xi^{-1})}, & \xi >1  \\
\\
\text{undefined} &  \text{otherwise}
\end{cases}
\end{equation}
For $\xi < 1$, we can still study the reliability of an EED, however, estimating the mean task execution time becomes implausible due to distribution tail behaviour. In fact, with unfinished tasks or tasks that take a long time to execute, estimating the EED's mean time for task execution becomes infeasible. Since we considered $\xi$ as a constant, the parameter $\alpha$ can be defined as the EED mean task execution rate (tasks/sec). The EED task execution rate depends mainly on three factors: 1) The computational capacity of the EED, 2) the utilization of the computational resources for the task (i.e., the allocated computational resources for the task), 3) The computational demand of the task. Herein, we define the task execution rate as the following:
\begin{equation}
\lambda = \frac{u C}{d},
\end{equation}
where $C$ is the computational capacity (cycles/sec) of the EED, $u$ is the utilization such that $u \in [0,1]$, and $d$ is the task demand (cycles/task). In this work, we consider the utilization variable as the available computational resources that are allocated from the EED to the task. In addition, it acts as an indicator of how much the EED is utilizing from its full computational capacity, where it abstracts many factors (e.g., the scheduling algorithm, the number of threads inside the CPU, etc.). 

However, considering a constant task execution rate for a task is impractical, since there are always other local tasks on the EED side, besides the lack of complete knowledge about the EED local computation behavior. Hence, the task execution rate can be one of the following:
\begin{enumerate}
   
    \item An RV, where the utilization $U$ and the task demand $D$ are RVs that follow some probability distributions (i.e., $U\sim f_U$ and $D\sim f_D$). Hence, the mean task execution rate is given by:
    \begin{equation}
        \alpha = C~\mathbb{E}[\lambda] =  C~ \mathbb{E}\left [\frac{U}{D} \right] .
    \end{equation}
    \item A function of time, i.e., $\lambda(t)$, where the utilization and the task demand changes deterministically with time. As such, the mean task execution rate at time $t$ is the function average and is given by: 
    \begin{equation}
     \alpha(t) =\frac{1}{t} \int_{0}^t \lambda(\tau) d\tau =  \frac{ C  }{t} \int_{0}^t \frac{u(\tau) }{d(\tau)} d\tau  .
    \end{equation}
    \item A stochastic process, where the utilization and the demand are time indexed RVs. The mean task execution rate at time $t$ is then given by:
    \begin{equation}
        \alpha(t) = C~ \mathbb{E}[\lambda(t)] =  C~ \mathbb{E}\left [\frac{U(t)}{D(t)} \right] .
    \end{equation}
\end{enumerate}

In this work, we will elaborate on the first case where there is uncertainty with known distribution in the task demand and the EED utilization of their computational resources, while we keep the doors open for future contributions for the other cases. 

Let $D$ be a uniform RV that follows the distribution $f_D$, where $f_D$ represents the task demand distribution with a range of $\left [ D_{\text{min}}, D_{\text{max}}\right ]$, with $D_{\text{min}}$ and $D_{\text{max}}$ denoting the minimum and maximum demands possible, respectively. 

The first moment or the average demand $D_{\text{m}}$ is given by: $D_{\text{m}} = \frac{D_{\text{max}} + D_{\text{min}}}{2}$. Thereafter, the reciprocal RV ,$D^{-1} = \frac{1}{D}$, follows an inverse uniform distribution, and its first moment is defined as:
\begin{equation}
\mathbb{E}[D^{-1}] = \frac{\log(D_{\text{max}}) - \log (D_{\text{min}})}{D_r} = \frac{\log \left( \frac{4 D_{\text{m}}}{2D_{\text{m}}- D_r} - 1 \right)}{D_r}
\end{equation}
where $D_r =  D_{\text{max}} - D_{\text{min}}$ is the demand range. Let $U$ be another RV that follows a uniform distribution $f_U$, where $f_U$ represents the utilization distribution with a range of $\left [ U_{\text{min}}, U_{\text{max}}\right ]$, with $U_{\text{min}}$ and $U_{\text{max}}$ denoting the minimum and maximum utilization possible, respectively, and an average of $U_{\text{m}} = \frac{U_{\text{max}} + U_{\text{min}}}{2}$. 

Since the demand and the utilization are non-negative RVs, then according to Melvin \cite{AlgebraofRandom}, the first moment of their division can be expressed as:

\begin{equation}
\mathbb{E}\left [\frac{U}{D} \right] = \mathbb{E}\left [U\right] \mathbb{E}\left [\frac{1}{D}\right],
\end{equation}
and hence, by substituting in Eq. (5), the mean task execution rate can be given by:
\begin{equation}
\alpha = \frac{ C ~ U_{\text{m}} ~  \log \left( \frac{4 D_{\text{m}}}{2D_{\text{m}}- D_r} - 1 \right)}{D_r}
\end{equation}
One can see that the mean task execution rate considering uniform distributions depends mainly on the average computational capacity utilization, the average, and the range of the task demand. Lastly, considering $\xi>1$, the reliability of an EED given a time deadline $t$ can be expressed as:

\begin{equation}
R(t) = 1- \left (1+ \frac{t~ C ~ U_{\text{m}} ~  \log \left( \frac{4 D_{\text{m}}}{2D_{\text{m}}- D_r} - 1 \right)}{\xi ~ D_r} \right)^{-\xi}
\end{equation}


\section{Modeling of Extreme Edge Systems Reliability}\label{sec:Sys_rel}
In this section, we study the reliability of EEC systems. 
We consider a simple EEC system that consists of one orchestrator and $n$ EEDs. We assume the case of distributed tasks only, where the orchestrator is responsible for distributing the task across the EEDs as sub-tasks, while the EEDs are responsible for performing the computation required for the sub-tasks. We define the reliability of an EEC system as the probability that the whole system accomplishes the distributed task on time. There are mainly two types of distributed tasks:  1) Tasks whose results depend only on a single completion, such that if only one EED finished its sub-task, the task result will be carried out, and 2) Tasks whose results depend on the completion of all the distributed sub-tasks.
Accordingly, we categorize EEC systems into two types, namely, series systems and parallel systems. 

\subsection{Series Systems}
In series systems, the total system reliability for distributing a task is defined for tasks whose result depends on the completion of \textbf{all} sub-tasks. Furthermore, such tasks also have two different types:
\begin{enumerate}
    \item Series non-Sequential (SNS) tasks, where each sub-task does not require any results from other sub-tasks. However, the system will wait until all the EEDs to finish their sub-tasks. In fact, it will wait for the EED that takes the longest time in order to carry out the result of the task. Examples of such tasks can be found in blockchain, when there exist $n$ validators, and all of them are required to validate a transaction. In such case, the slowest node dictates the total validation time. 

    \item Series Sequential (SS) tasks, where the result of each sub-task depends on the results of other sub-tasks. An example of such tasks is Distributed ML Inference \cite{baccour2020distprivacy}, where a trained ML model is divided into multiple segments, and each segment is sent to an EED to perform the required segment computation. Each segment in the model requires the output from the previous segments. As a result, the ML model output can only be available if and only all the EEDs performed the segments computation successfully in order. 
    
\end{enumerate}

Let $T_1, T_2,...,T_n$ be independent RVs which represent the time taken by the EEDs to finish their sub-tasks. Let $T_{\text{max}} = \max(T_1,T_2,...,T_n)$, which represents the time of the EED that takes the maximum time until it finishes its task. Therein, the probability that the EEC SNS system finishes the task by a time deadline of $t$ is simply as follows:
\begin{equation}
\begin{split}
 &R_s(t) := P(T_{\text{max}}\leq t) = P(\max(T_1,T_2,...,T_n)\leq t)  \\ 
 &=\prod^n_{i=1} F_i(t)
\end{split}
\end{equation}
where $F_i$ is the CDF function from Eq (2).

Afterwards, let $T_1, T_2,...,T_n$ be the ordered time taken by $n$ EEDs to finish their ordered sub-tasks. Therein, the probability of the EEC SS system finishing the task on time $t$ is as follows:
\begin{equation}
\begin{split}
 & R_s(t):= P\big((T_1,T_2,...,T_n)\leq t\big)  \\
 &=\prod^n_{i=1} F_i \left( \max\left(t-\sum^{i-1}_{j=1} T_j,0\right) \right)
\end{split}
\end{equation}
where the term  $\sum^{i-1}_{j=1} T_j$ represents the time taken by previous EEDs to finish their sub-tasks. The max function indicates that if the time taken by the previous EEDs exceeds the time deadline $t$, then the system fails to finish the task, and hence, has a reliability of 0.

\subsection{Parallel Systems}
In parallel systems, the total system reliability for distributing a task is defined for tasks whose result depends only on a single sub-task completion out of all sub-tasks. In other words, the task result can be carried out as soon as the first EED finishes its sub-task. An example of such tasks is Federated Learning, where a global ML model needs to be trained on the local datasets of the EEDs. The global model is distributed across the EEDs as a sub-task, and each EEDs perform the computation required for the ML model training individually, and lastly, the orchestrator consolidates all the trained models into one global model. In fact, a trained global model can be available even if only one EEDs trained the model, even though the model quality might be poor. 

Let $T_1, T_2,...,T_n$ be the time taken by $n$ EEDs to finish their individual sub-tasks. Let $T_{\text{min}} = \min(T_1,T_2,...,T_n)$, which represents the time of the EED that takes the minimum time to finish its sub-task, then the system reliability is the probability of that EED finishing on time $t$ and is given by:
\begin{equation}
\begin{split}
 &P(T_{\text{min}}\leq t) = P(\min(T_1,T_2,...,T_n)\leq t)  \\ &=1-\prod^n_{i=1} \left (1-F_i(t)\right)
\end{split}
\end{equation}

\section{Simulation Results}\label{sec:sim}
\begin{figure}[t]
    \centering
    \includegraphics[scale=0.67]{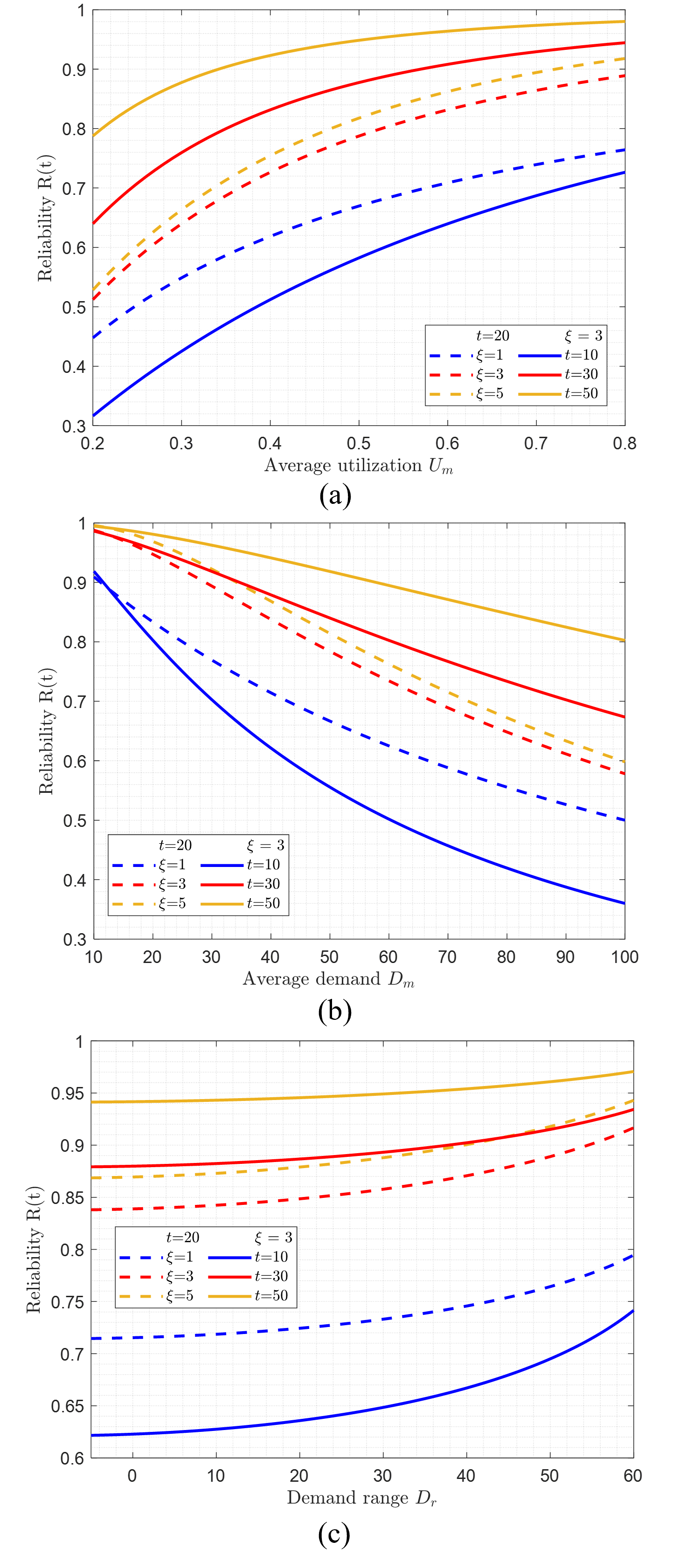}
    \caption{The reliability behaviour with different time deadlines and $\xi$'s with respect to (a) average resource utilization (b) average task demand (c) demand range.}
    \label{fig:rel_EED}
\end{figure}

In this section, we first demonstrate the reliability behaviour of the EEDs considering the GDP distribution. In Figure \ref{fig:rel_EED},  we show how the reliability changes with respect to average computational resource utilization, average task demand, and the range for the task demand, considering different time deadlines and values for the constant $\xi$. Figure \ref{fig:rel_EED} (a) depicts the reliability against the average resource utilization considering $C = 1$ GHz, $D_r = 2 \times 10^9 $ cycles and $D_m =5 \times 10^9 cycles$. It can be observed that the reliability increases as we increase the utilization of the computational resources. Intuitively, more resources allow higher task execution rate, and hence, higher reliability. Moreover, we can notice that as we increase the parameter $\xi$, the EED can achieve better reliability behaviour. Similarly, as we increase the time deadline of the task, the EED can have more time to execute the task and achieve better reliability. In Figure \ref{fig:rel_EED} (b), the reliability versus the average task demand is shown considering $U_{\text{m}} = 0.7$. It can be seen that as the task demand increases, the reliability decreases. In fact, the increase of the task demands while fixing the resources' utilization imposes more load on the EED and lowers the task execution rate, and therefore, the EED becomes less reliable given the same time deadline. In a similar behavior, as the constant $\xi$ and the time deadline $t$ increase, the EED achieves better reliability behaviour overall. Lastly, the reliability versus the demand range is shown in Figure \ref{fig:rel_EED} (c). Interestingly, as the task demand range increases, the reliability slightly increases. As a matter of fact, with a wider range of the computational demand, fewer demands become more likely to occur, allowing the EEDs to have a better task execution rate. 

\begin{figure}[ht!]
    \centering
    \includegraphics[scale=0.67]{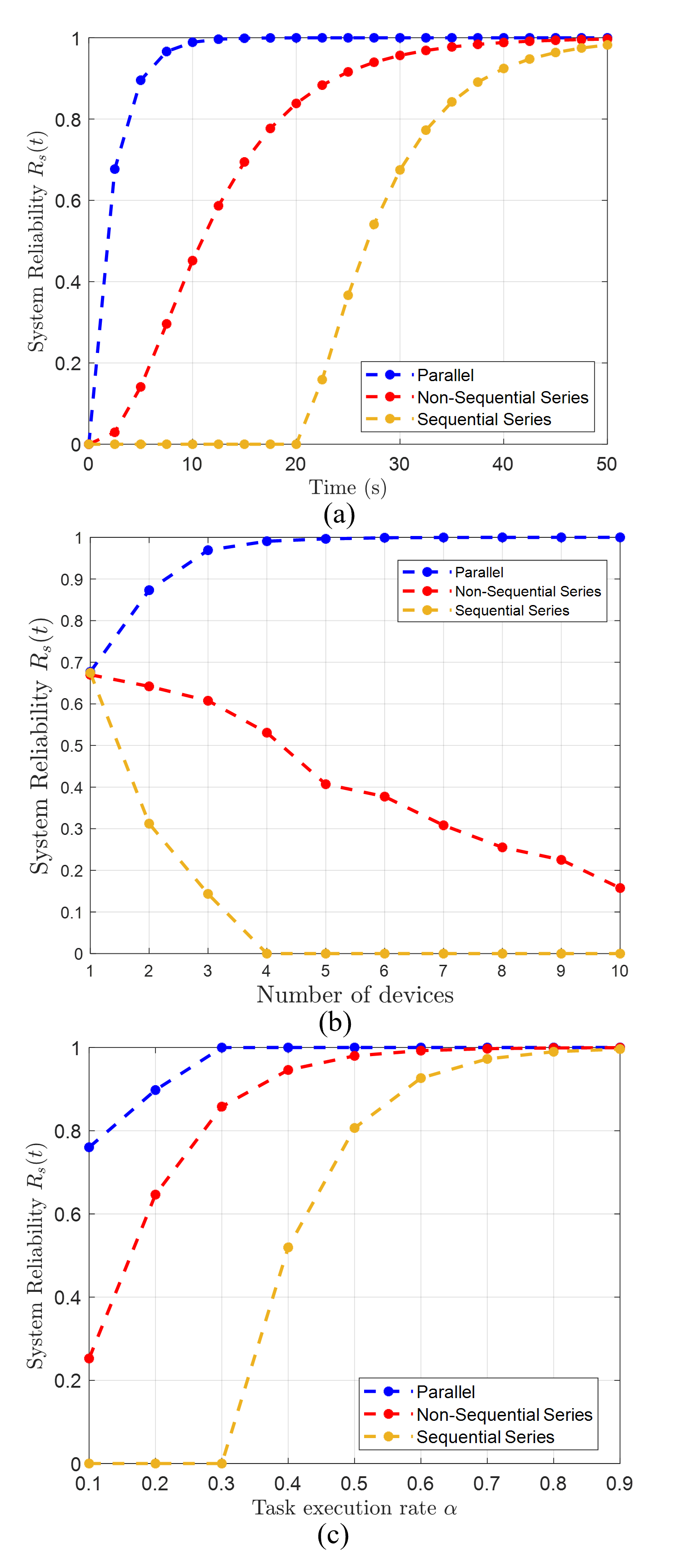}
    \caption{The reliability of EEC systems while varying (a) time deadline (b) number of devices (c) task execution rate}
    \label{fig:Sys_rel}
\end{figure}

Afterward, we show the EEC systems reliability for the parallel, SNS, and SS systems, while varying the time deadline, the number of devices within the system, and the abstract task execution rate of the EEDs in Figure \ref{fig:Sys_rel} with the assumption of $\xi^{-1}=0$. Figure \ref{fig:Sys_rel} (a) depicts the system reliability against the time deadline of the task. It can be noticed that the parallel system has the best reliability behaviour and converges to its maximum faster than the other systems. In addition, one can notice that the SS systems have reliability of 0 for shorter time deadlines. Indeed, with short time deadlines, the time taken by EEDs to execute their sub-tasks exceeds the deadline, and as a consequence, that task has not been accomplished on time. Subsequently, the system reliability as we increase the number of EEDs in the system is depicted in Figure \ref{fig:Sys_rel} (b). We can see that with more EEDs in the system, the reliability of the parallel system increases, whereas the series system decreases. As a matter of fact, increasing the number of EEDs in parallel systems increase the chances of that task being executed, since only one EED is needed to carry out the task result. On the contrary, increasing the number of EEDs in a series system lowers the chances of that task being executed, since the result is dependent on all the EEDs' sub-task execution. In other words, more dependencies with uncertainty decrease the reliability of the overall system. In addition, the SS systems reliability goes to 0 after a certain number of EEDs in the system. Basically, with more EEDs in the SS system, the total time taken by the EEDs to finish the task increases, and once the total time exceeds the deadline, the task cannot be finished on time. Last but not least, the system reliability as we increase the EEDs average task execution rate in the system is depicted shown Figure \ref{fig:Sys_rel} (c). Intuitively, if the task execution rate increases the individual EEDs reliability, then the total system reliability also increases. Also, for low task execution rates, the SS systems have also 0 reliability, because with low execution rates the EEDs also cannot finish the task on time.

\section{Discussions and Future Directions}\label{sec:discussion}

The reliability of EEDs can be used as a metric for various applications such as client selection or scheduling, where the aim can be to select or schedule the most reliable devices to execute the task. Moreover, it can be used in designing incentive mechanisms, such that EEDs that are more reliable can be paid more or cost more. In addition, the system trade-offs in terms of reliability and other metrics such as cost can be further studied. For instance, in parallel systems, recruiting more EEDs increases the system reliability, however, it costs more to recruit more EEDs. The following question arises here: What is the optimal number of devices to recruit such that we can maximize the system reliability and minimize the recruitment costs? Costs can be an abstraction of other metrics such as energy consumption, latency, monetary pricing, etc. 

Throughout this work, we considered the GDP distribution for the task times at the EEDs. Even though there exist other plausible distributions such as Half-Caushy or Half-Log Normal, the GDP distribution is the strongest candidate as it is a generalization of multiple distributions, in addition to its ability to be tweaked such that it can fit any exponentially looking data. Moreover, we assumed uniform distributions for the resources' utilization and the task computational demand for ease of analysis. In fact, for the resources' utilization, such assumption is valid in EEC systems, where EEDs can estimate the range of how much resources they can utilize for that task, guaranteeing a minimum and maximum value for the utilization, where the utilization could be anywhere within the estimated range.

As for future directions, analyzing the EEDs reliability for streaming services is of great importance. Reliability for streaming services refers to the ability of the EED of offering an uninterruptible and seamless service to the end users continuously. The reliability should be analyzed at each point in time, and is not associated only with a time deadline. In addition, other realistic distributions for the task or service computational demands can be also studied to provide a more realistic reliability model. Furthermore, the other cases for the task execution rate (e.g., a function of time or a stochastic process) should also be analyzed to accommodate different types of tasks and services in the model.
Finally, since the aim is to provide a general and realistic reliability model for all EEC systems, a further study that takes into account multiple orchestrators with multiple tasks or services and the queuing analysis of the system is of utmost importance.

\section{Conclusion}\label{sec:conc}
In this work, we studied the EEDs randomness from the computational perspective, and how reliable is an EED in terms of executing the tasks on time. Specifically, we modeled the reliability of the EEDs while taking into account the probabilistic nature of EED's computational resources, along with the tasks' computational demand. Furthermore, we studied the EEC systems' reliability while considering different types of tasks in EEC systems. Lastly, we carried out simulation results to show the EEDs reliability behavior in different scenarios, in addition to the EEC systems' reliability.  
\section*{Acknowledgement}
This work was made possible by NPRP grant \# NPRP13S-0205-200265 from the Qatar National Research Fund (a member of Qatar Foundation). This work was also supported by a grant from the Natural Sciences
and Engineering Research Council of Canada (NSERC)
under grant number: [ALLRP 549919-20]. The findings achieved herein are solely the responsibility of the authors.  

\bibliographystyle{IEEEtran}
\bibliography{references.bib}

\end{document}